\documentclass{PoS}

\title{Key Science Observations of AGNs with KaVA Array}

\ShortTitle{Key Science Observations of AGNs with KaVA}

\author{\speaker{Motoki Kino}
\\
        Korea Astronomy and Space Science Institute (KASI), 
        776, Daedeokdae-ro, Yuseong-gu, Daejeon, Republic of Korea 305-348\\
        E-mail: \email{kino@kasi.re.kr}}

\author{Kotaro Niinuma\\
Graduate School of Science and Engineering, 
Yamaguchi University, 
Yoshida 1677-1, Yamaguchi, Yamaguchi 753-8512,
Japan\\}

\author{Guang-Yao Zhao,  Bong Won Sohn\\
         Korea Astronomy and Space Science Institute (KASI), 
        776, Daedeokdae-ro, Yuseong-gu, Daejeon, Republic of Korea 305-348\\}

\author{KaVA AGN Sub Working Group\\}
        
\abstract{
KaVA (KVN and VERA Array) is a new combined VLBI array 
with KVN (Korean VLBI Network) and VERA (VLBI Exploration of Radio Astrometry).
First, 
we briefly review the imaging capabilities of KaVA array
which actually achieves more than three times better 
dynamic range than that achieved by VERA alone.
The KaVA images clearly show detailed structures of
extended radio jets in AGNs. 
Next, we represent  the key science program 
to be led by KaVA AGN sub working group.  
 We will conduct the monitoring observations of Sgr A* and M87 
 because of the largeness of their 
 central super-massive black hole angular sizes. 
 The main science goals of the program are 
 (i) testing magnetically-driven-jet paradigm by mapping velocity fields of the M87 jet, and
 (ii) obtaining tight constraints on physical properties of radio emitting region in Sgr~A*.}

\FullConference{12th European VLBI Network Symposium and Users Meeting,\\
		7-10 October 2014\\
		Cagliari, Italy}

\begin{document}

\section{Introduction}

Active galactic nuclei (AGN) occasionally produce 
powerful and collimated
jets of magnetized relativistic particles which 
can extend beyond galactic scales, and  
have an impact on galaxy evolutions.
Although 
magnetic driven scenario of relativistic jets
is widely discussed
(e.g., 
\cite{BZ77};
\cite{M06}),
there is no conclusive observational evidence
to prove it. 
In order to explore jet formation processes
by a central engine composed of a supermassive black hole
and accreting matter onto it, 
Very Long Baseline Interferometer (VLBI) is 
one of the most powerful tools because
they can probe innermost regions of relativistic jets with 
its high spatial resolution.

Recently, a new VLBI facility, named KaVA, consisting of Korean
VLBI network (KVN) and VLBI Exploration of Radio
Astrometry (VERA) has been constructed in East Asia
region (http://veraserver.mtk.nao.ac.jp/). 
KVN is the first VLBI array dedicated to the
mm-wavelength radio observations in East Asia operated
by Korean Astronomy and Space Science Institute
(KASI) (\cite{LPB14})
 (http://kava.kasi.re.kr/kava main.php). 
 KVN consists
of three 21-m-diameter radio telescopes: one in Seoul,
one in Ulsan, and one on Jeju Island, Korea.
%%
%In each, four different frequency band receivers are installed 
%(22, 43, 86, and 129 GHz).
%%
In this proceedings, 
AGN  Key Science Project (KSP) with KaVA, i.e., 
monitoring M87 and Sgr A* at 23~GHz and 43~GHz
is summarized.

\section{Imaging Capability of KaVA}

Here we briefly review imaging capability of KaVA.
In radio interferometers, the detection limit 
(equivalent to thermal noise level) 
of images is given by
$\sigma_{\rm th} 
= \frac{2k_{\rm B}T_{\rm sys}}
{A_{\rm eff}\eta_{\rm q}\sqrt{N_{\rm ant}(N_{\rm ant}-1)BW t_{\rm int}}}$
where 
$A_{\rm eff}$,
$\eta_{\rm q}$,
$T_{\rm sys}$, 
$N_{\rm ant}$,
$BW$,
$t_{\rm int}$, are
the efficient aperture area of the antennas, 
the quantization loss factor,
the system temperature,
the number of antennas,
bandwidth, and
the total integration time, respectively.
When using natural weighting,  a typical thermal
noise for KaVA observation of M87 at 23~GHz 
with $BW=32$~MHz is 
$\sigma_{\rm th}\approx 0.7~{\rm mJy~beam^{-1}}$
and the corresponding signal-to-noise ratio (SNR)  is $\sim 1400$
(see details for \cite{NLK15}).

\begin{figure*}[t]
\centering
\includegraphics[width=60mm]{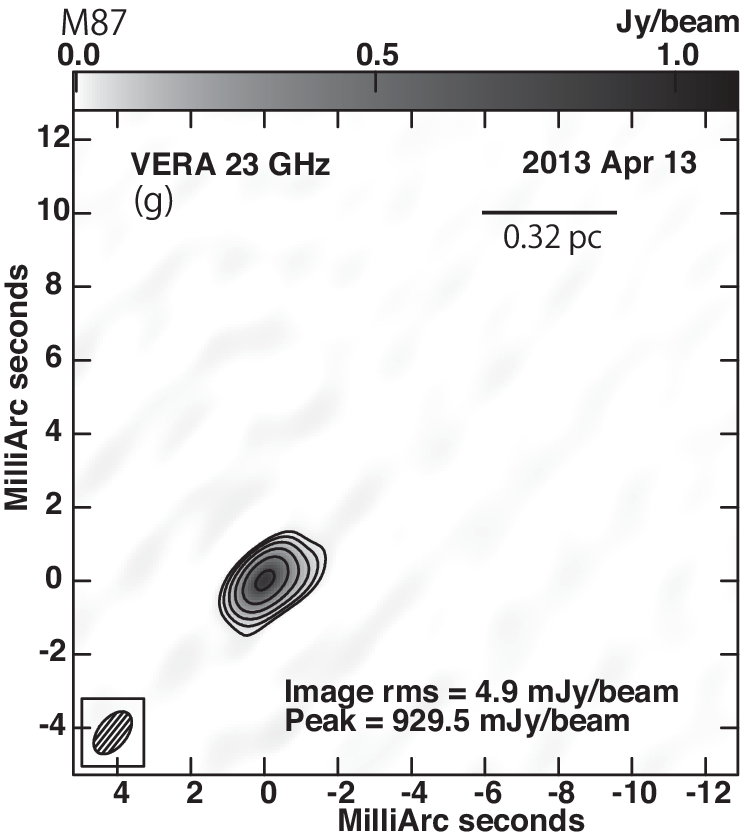}
\includegraphics[width=60mm]{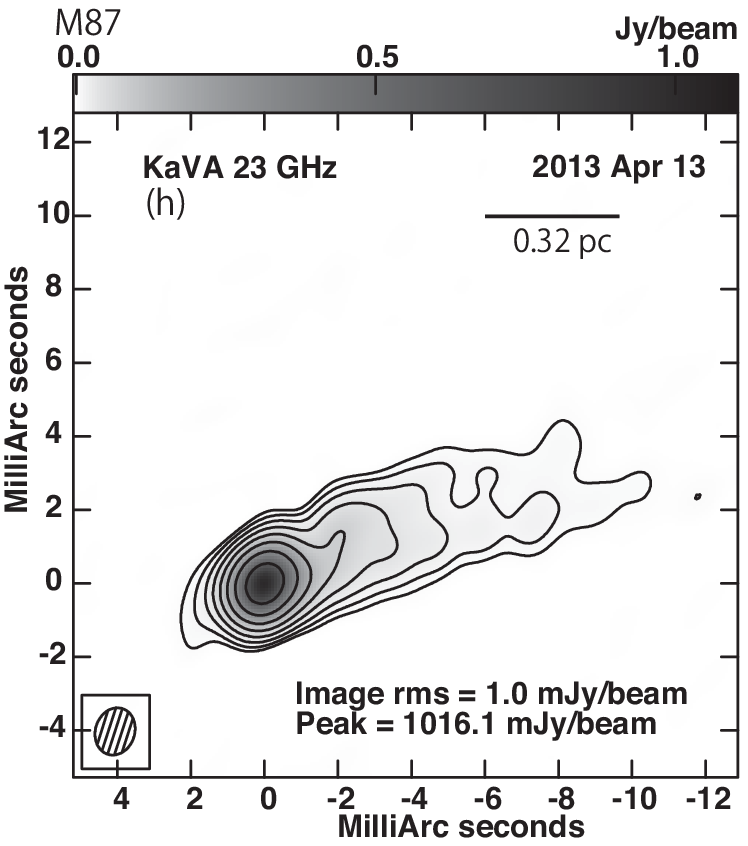}
\caption{The comparison of VERA and KaVA images of M87 at 23~GHz. 
%The higher dynamic range of KaVA enables us 
%to obtain the extended structure of the  M87 jet up to 10~mas scale.
\label{fig:m87-32M}}
\vspace{5mm} %% add extra space ONLY when figures/tables are "colliding"!
\end{figure*}

In Figure \ref{fig:m87-32M}, we show
the comparison of VERA and KaVA images of M87 at 23~GHz. 
The higher dynamic range of KaVA enables us 
to obtain the extended structure of the  M87 jet up to 10~mas scale.
Qualities of VLBI images tend to be limited by their dynamic ranges
rather than SNR. The dynamic range of VLBI images can be given by
$\frac{I_{p}}{\sigma_{\rm im} }=
 \frac{\sqrt{M_{\rm scan}}\sqrt{N_{\rm ant}(N_{\rm ant}-1)}}{max[\epsilon, \Delta \phi]}$
where $M_{\rm scan}$,
$\epsilon$, and
$\Delta \phi$ are the number of scans for each observation, and
degrees of amplitude and phase errors, respectively. 
%%%
The dynamic range of KaVA image in Figure \ref{fig:m87-32M}
reaches $\sim 1000$ and it is more than three times  
better than that achieved by VERA alone (\cite{NLK15}).

\section{Key Science Project\label{sec:sections}}

\subsection{M87}

M87, a nearby giant radio galaxy located
at a distance of 16.7~Mpc, hosts one of the most 
massive super-massinve black holes with it mass of
 $6\times 10^{9}~M_{\rm sun}$.
Thanks to its proximity and the largeness of its central black hole,
M87 is well known for being the best source for
imaging the innermost  part of the jet base
(e.g., \cite{HDK11}).
According to the leading scenario of jet formation model, 
a jet is thought to be powered by a central engine
in a highly magnetized state, and accelerated 
via conversion of magnetic energy into kinetic one. 
Relativistic magnetohydrodynamic models have
suggested that  jets are gradually accelerated 
on a scale which can be well observed by VLBI 
in the case of M87 (e.g., \cite{M06}). 
Hence, it is possible to constrain such  models 
by comparing the model-predicted velocity fields  and 
observed one.
Indeed, mapping the apparent velocities of the M87 jet
has been explored in previous work \cite{Asada14}. 
However, 
the reported apparent velocities at 1-10~mas from the central engine 
are significantly different  and it is controversial. 
The aim of this KSP is measuring the actual velocity field in
the M87 jet to test magnetically-driven jet paradigm 
with sufficiently short interval to avoid 
possible component mis-identifications.

\subsection{Sagittarius~A*}

\begin{figure*}[t]
\centering
\includegraphics[angle=-90,width=60mm]{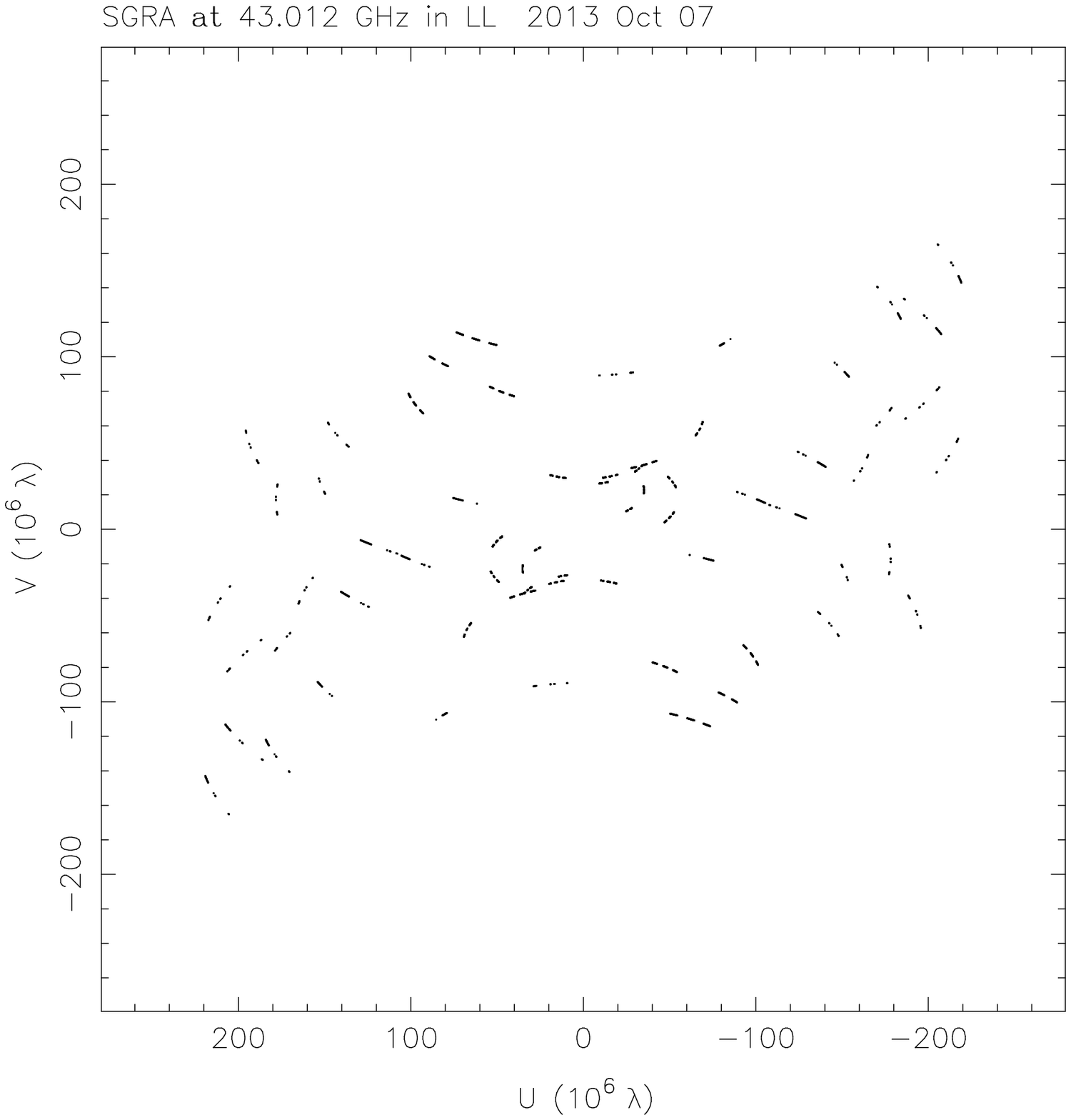}
\hspace{2cm}
\includegraphics[angle=-90,width=50mm]{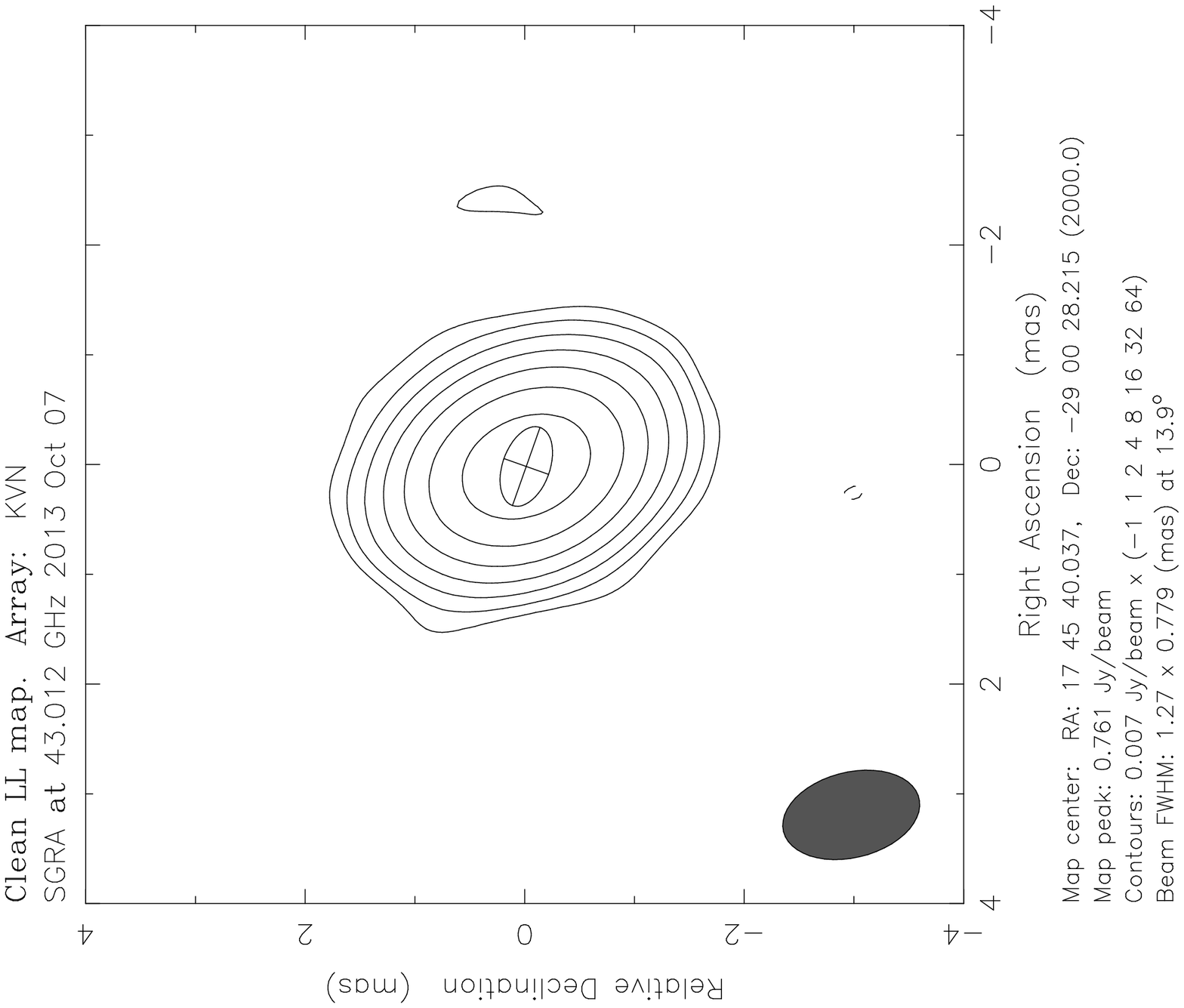}
\caption{
Left: The $u, v$ coverage of KaVA observation of Sgr A* at 7~mm 
performed on Oct 7th 2013.
Right:
Corresponding KaVA  image of Sgr A* 
with the best-fit elliptical Gaussian model. 
%%
%The position angle, 
%the sizes of major and minor axises, 
%peak intensity, 
%total flux, are ............, respectively.
%%
}
\label{fig:sgra-kava}
\vspace{5mm} %% add extra space ONLY when figures/tables are "colliding"!
\end{figure*}

The center of Milky Way hosts Sgr~A*, a massive black hole
 with the largest angular size, hence it is one of the best laboratories to explore ultimate vicinity of black holes
(e.g., \cite{DWR08}).
During the past VLBI monitoring of Sgr~A* at 43~GHz, the flare phenomena 
was found in 2007 \cite{ATH13}. 
Interestingly, the size of the major axis remained the same while the flux 
increased about 2~Jy. This is the first report for VLBI scale flare and we do not know how 
frequently it happens.
In Figure~\ref{fig:sgra-kava}, we present the $u,v$ coverage and 
the preliminary image of Sgr~A* at 43~GHz with the bandwidth of 256~MHz.
We emphasize that 
KaVA can achieve a good performance for Sgr~A* observations, since it
contains more short baselines than other VLBI arrays. 
Short baselines in KaVA provide
more effective sampling of the visibilities of Sgr~A* than VLBA 
and it enables us to get better measurements of the source size.
%%%
Accurate measurements of size and flux of Sgr A* by KaVA at 43~GHz
will enable us to give  tight constraints on physical quantities in Sgr~A*
such as magnetic field strength 
since these quantities have strong dependences on the source size \cite{K14}.

%%% FIGURE %%%%%%%%%%%%%%%%%%%%%%%%%%%%%%%%%%%%%%%%%%
%%%%%%%%%%%%%%%%%%%%%%%%%%%%%%%%%%%%%%%%%%%%%%%%%%%%%
%%% FIGURE %%%%%%%%%%%%%%%%%%%%%%%%%%%%%%%%%%%%%%%%%%%%%%
%%%%%%%%%%%%%%%%%%%%%%%%%%%%%%%%%%%%%%%%%%%%%%%%%%%

\end{document}